\documentclass[aps,prc,amsfonts,preprintnumbers,superscriptaddress,showpacs,nofootinbib]{revtex4}

\usepackage{graphics}
\usepackage{psfrag}
\usepackage{epsfig}
\usepackage{epsf}
\usepackage{float}

\newcommand{\fet}[1]{\mbox{\boldmath $#1$}}

\newcommand{\beq}{\begin{equation}}
\newcommand{\eeq}{\end{equation}}
\newcommand{\beqa}{\begin{eqnarray}}
\newcommand{\eeqa}{\end{eqnarray}}
\newcommand{\nn}{\nonumber \\ }

\newcommand{\krig}[1]{\stackrel{\circ}{#1}}

\begin{document}

\preprint{FZJ-IKP-TH-2007-34}
\preprint{HISKP-TH-07/28}

\title{$\Delta$-excitations and the three-nucleon force}

\author{E.~Epelbaum}
\email[]{Email: e.epelbaum@fz-juelich.de}
\affiliation{Forschungszentrum J\"ulich, Institut f\"ur Kernphysik 
(Theorie), D-52425 J\"ulich, Germany}
\affiliation{Universit\"at Bonn, Helmholtz-Institut f{\"u}r
  Strahlen- und Kernphysik (Theorie), D-53115 Bonn, Germany}
\author{H.~Krebs}
\email[]{Email: hkrebs@itkp.uni-bonn.de}
\affiliation{Universit\"at Bonn, Helmholtz-Institut f{\"u}r
  Strahlen- und Kernphysik (Theorie), D-53115 Bonn, Germany}
\author{Ulf-G.~Mei{\ss}ner}
\email[]{Email: meissner@itkp.uni-bonn.de}
\homepage[]{URL: www.itkp.uni-bonn.de/~meissner/}
\affiliation{Universit\"at Bonn, Helmholtz-Institut f{\"u}r
  Strahlen- und Kernphysik (Theorie), D-53115 Bonn, Germany}
\affiliation{Forschungszentrum J\"ulich, Institut f\"ur Kernphysik 
(Theorie), D-52425 J\"ulich, Germany}
\date{\today}

\begin{abstract}
We study the three-nucleon force in chiral
effective field theory with explicit $\Delta$-resonance degrees of freedom. 
We show that up to next-to-next-to-leading order, the only contribution
to the isospin symmetric three-nucleon force involving the spin-3/2 degrees of freedom is 
given by the two-pion-exchange diagram with an intermediate delta, frequently
called the Fujita-Miyazawa force. We also analyze the leading isospin-breaking
corrections due to the delta. For that, we give the first  quantitative analysis of the
delta quartet mass splittings in chiral effective field theory including 
the leading electromagnetic corrections. The
charge-symmetry breaking three-nucleon force due to an intermediate delta excitation is small,
of the order of a few keV.
\end{abstract}


\maketitle

\vspace{-0.2cm}

\section{Introduction and summary}
\def\theequation{\arabic{section}.\arabic{equation}}
\setcounter{equation}{0}
\label{sec:intro}

The importance of the $\Delta$-excitation in the three-nucleon force (3NF)
has been already realized fifty years ago by Fujita and Miyazawa in their
seminal work \cite{FuMi}. Their paper  has been the seed for many meson-theoretical 
approaches to the 3NF like the families of Tucson-Melbourne \cite{TM1,TM2}, 
Brazilian \cite{Brazil} or Urbana-Illinois \cite{Urbana1,Urbana2} 
3NFs, see the review article \cite{GloeckleRept} and the 
recent general introduction \cite{KalantarNayestanaki:2007zi}.
Nowadays, the appropriate tool to 
analyze the forces between nucleons is chiral effective field theory, a
program started by Weinberg \cite{Weinberg:1991um}. There has been quite
a sizeable body of further work on the structure of 3NFs within the framework
of EFT, with appropriate references given below. Still, in the EFT
with explicit deltas most investigations so far have considered the effects
based on the leading pion-nucleon-delta coupling. In this work,
we want to go one step further. In the two-nucleon system, we had already
considered the effects of subleading $\pi N \Delta$ couplings on the
description of the peripheral phase shifts \cite{Krebs:2007rh}. Such 
effects appear at  next-to-next-to-leading order (NNLO) in the Weinberg counting.
It is therefore appropriate to extend these considerations to the 3NF,
more precisely to the $\Delta$-contributions to the
isospin-symmetric 3NF at NNLO and the leading isospin-breaking corrections to
the 3NF. An important ingredient in latter type of forces stems from the
mass splittings in the quartet of $\Delta$-states ($\Delta^{++}, \Delta^+, 
\Delta^0, \Delta^-$). These splittings receive contributions from the
strong and electromagnetic interactions, thus one needs to consider an
appropriate extension of the power counting including the explicit soft and hard
photon effects.

The pertinent results of this investigation can be summarized as follows:
\begin{itemize}
\item[i)] We have evaluated the $\Delta$-contributions to the three-nucleon 
force up to NNLO and shown that the only non-vanishing topology is the
two-pion exchange diagram with an intermediate delta resonance, commonly 
called the Fujita-Miyazawa force. This implies that the leading
contributions to the $1\pi$-$4N$-contact and the $6N$-nucleon contact
topologies are not saturated by the $\Delta$. 
\item[ii)] We have compared the $2\pi$-exchange 3NF in the EFT with and
without explicit $\Delta$ degrees of freedom. We show that these
representations lead to comparable results for the strength of the 3NF 
if the same data basis for
pion-nucleon scattering is used to pin down the LECs in the two versions
of the theory, see table~\ref{tab:abd}.
\item[iii)] We have analyzed for the first time the mass splitting 
in the delta quartet from strong and electromagnetic contributions 
with EFT up-to-and-including
second chiral order. Note that chiral extrapolation formulae for the strong splitting
have already been given earlier in Ref.~\cite{Tiburzi:2005na}.
The delta mass splittings are parameterized in terms
of two independent parameters, cf. Eq.~(\ref{deltas}). These can only
be determined with large uncertainties since the available information
on the various delta masses is fairly scarce and uncertain, 
see tables~\ref{tab2a},\ref{tab2}.
\item[iv)] We have shown how the proton-neutron mass difference can be
eliminated from the delta-full EFT by field redefinitions, extending the 
method developed in \cite{Friar:2004ca}. This facilitates the calculation
of the isospin-breaking effects to the 3NF considerably.
\item[v)] We have worked out the leading isospin-breaking contributions to
the 3NF due to an intermediate $\Delta$-excitation, see Fig.~\ref{fig1}.
The leading charge-symmetry conserving and charge-symmetry breaking
contributions to the 3NF are given in Eq.~(\ref{res_mom}) in momentum
space. We also give the coordinate space representation. We estimate the
contribution from the  charge-symmetry breaking force to the 3N binding energy
to be of the order of a few keV.
\end{itemize}

The manuscript is organized as follows. In sec.~\ref{sec:nnlo} we investigate
the delta contributions to the 3NF up-to-and-including next-to-next-to-leading
order in the so-called small scale expansion (SSE) \cite{Hemmert:1997ye}. The
calculation of the leading
isospin-violating contributions is presented in sec.~\ref{sec:isospin}.
As a first step, we analyze the mass splittings within the delta quartet
and calculate the strong and electromagnetic contributions to the various
particles, see sec.~\ref{sec:deltamN}. In sec.~\ref{sec:Nmass} we discuss
field redefinitions to eliminate the proton-to-neutron mass shift from the
effective Lagrangian which considerably simplifies the calculation of the isospin-breaking
effects. The isospin-breaking 3NFs due to explicit deltas are 
then worked out in sec.~\ref{sec:3NFisov}.

\section{$\Delta$-contributions to the three-nucleon force up to NNLO}
\def\theequation{\arabic{section}.\arabic{equation}}
\setcounter{equation}{0}
\label{sec:nnlo}

Our calculations are based on Weinberg's power counting \cite{Weinberg:1991um}
utilizing the small scale expansion \cite{Hemmert:1997ye}. In this
framework, irreducible diagrams with two or more nucleons which give rise to
the nuclear forces are ordered according to the power $\nu$ of
the expansion parameter $Q/\Lambda $, where $Q$ collectively denotes
small pion four-momenta, the
pion mass, baryon three-momenta and the nucleon-delta mass splitting and 
$\Lambda$ is the pertinent hard scale.  For an irreducible $N$-nucleon
diagram, the power $\nu$ is given by:
\beq
\label{powc}
\nu = -2 + 2 N + 2 (L -C) + \sum_i V_i \Delta_i\,.
\eeq
Here, $L$, $C$ and  $V_i$ refer to the number of loops, separately connected
pieces and  vertices of type $i$, respectively. Further, the vertex dimension
$\Delta_i$ is given by
\begin{equation}
\label{chirdim}
\Delta_i = d_i + \frac{1}{2} b_i - 2\;,
\end{equation}
where  $b_i$ is the number of baryon field operators and $d_i$ is the number
of derivative, insertions of $M_\pi$ and/or the delta-nucleon mass splitting,
$\Delta \equiv m_{\Delta} - 
m_N$.\footnote{Notice that we use the symbol $\Delta$
  for both the spin-3/2 field and the $N\Delta$ mass splitting. It is,
  however, always evident from the context what is meant.}

For the calculation of the isospin-conserving three-nucleon force (3NF) up to NNLO
we use the effective chiral Lagrangian already given in \cite{Krebs:2007rh}. 
The only additional terms that need to be considered are the leading-order
$NN\rightarrow N\Delta$ and $N\Delta\rightarrow NN$ contact interactions
\beq
\label{cont}
\bar{T}_i^\mu N\bar{N}S_\mu\tau^i N + \mbox{h.~c.}\,,
\eeq
where $N$ denotes the large component of the nucleon field, $T_i^\mu$ is 
the large component of the delta field,
with $i$ an isospin and $\mu$ a Lorentz index. Furthermore, $\tau^i$ and
$S_\mu$ are Pauli isospin matrices and $S_\mu$ denotes the covariant spin
vector. For more details on the notation and the effective Lagrangian the
reader is referred to \cite{Fettes:2000bb}. The contact interactions in
Eq.~(\ref{cont}) were already considered by van Kolck \cite{vanKolck:1994yi}
who worked out the corresponding contributions to the 3NF. 
As pointed out in Ref.~\cite{Epelbaum:2005pn},  matrix elements of the
resulting 3NFs  between antisymmetrized $|NNN \rangle$ states vanish, see also
\cite{Epelbaum:2002vt} for a related discussion. The reason for vanishing of
these 3NF contributions can be understood already at the level of
the effective Lagrangian. To that aim, let us rewrite Eq.~(\ref{cont}) including  
the  spin and isospin $3/2$ projectors $P_{\mu \nu}^{3/2}$ and $\xi_{i
  j}^{3/2}$  explicitly:
\beq
\bar{T}_i^\mu N\bar{N}S_\mu\tau^i N=\bar{T}_j^\nu\xi_{j i}^{3/2}P_{\nu
  \mu}^{3/2}  N\bar{N}S^\mu\tau^i N,
\eeq
with 
\beq
\xi_{i j}^{3/2}=\frac{2}{3}\delta_{i j}-\frac{i}{3}\epsilon_{i j
  k}\tau^k,\quad
P_{\mu \nu}^{3/2}=g_{\mu \nu}-v_\mu v_\nu -\frac{4}{1-d}S_\mu S_\nu \,.
\eeq
Here, $v_\mu$ denotes the baryon four-velocity and $d$ the number
of space-time dimensions.  In four dimensions with
$v=(1,0,0,0)$, the spin $3/2$ projector reduces to
\beq
P_{i j}^{3/2}=-\left[\frac{2}{3}\delta_{i j}-\frac{i}{3}\epsilon_{i j k}
\sigma^k\right]\,,
\eeq
with $\sigma^k$ denoting the Pauli spin matrices.
It is straightforward to see that the antisymmetrized Feynman rules for the
particular contact interactions vanish:
\beq
\label{temp1}
\xi_{i j}^{3/2}(1)P_{k l}^{3/2}(1)\tau_2^j\sigma_2^l {\cal A}_{1 2}=
\xi_{i j}^{3/2}(2)P_{k l}^{3/2}(2)\tau_1^j\sigma_1^l {\cal A}_{1 2}=0\,,
\eeq
where the subscripts of the Pauli spin and isospin
matrices refer to the nucleon labels and 
the antisymmetrization operator for momentum-independent interactions 
${\cal A}_{1 2}$ has the form 
\beq
{\cal A}_{1 2}=\frac{1-P_{1 2}}{2},\quad {\rm with} \quad
P_{1 2}=\frac{1+\vec{\sigma}_1\cdot\vec{\sigma}_2}{2}
\frac{1+\fet{\tau}_1\cdot\fet{\tau}_2}{2}\,.
\eeq
The labeled projector operators $\xi_{i j}^{3/2}(X) $ and $P_{i j}^{3/2}(X)$
in Eq.~(\ref{temp1}) are defined according to
\beq
\xi_{i j}^{3/2}(X)=\frac{2}{3}\delta_{i j}-\frac{i}{3}\epsilon_{i j
  k}\tau_X^k,\quad 
P_{i j}^{3/2}(X)=-\left[\frac{2}{3}\delta_{i j}-\frac{i}{3}\epsilon_{i j k}
\sigma_X^k\right],\quad X\in\{1,2\}\,.
\eeq 

\begin{figure}[tb]
\vskip 1 true cm
\includegraphics[width=11.0cm,keepaspectratio,angle=0,clip]{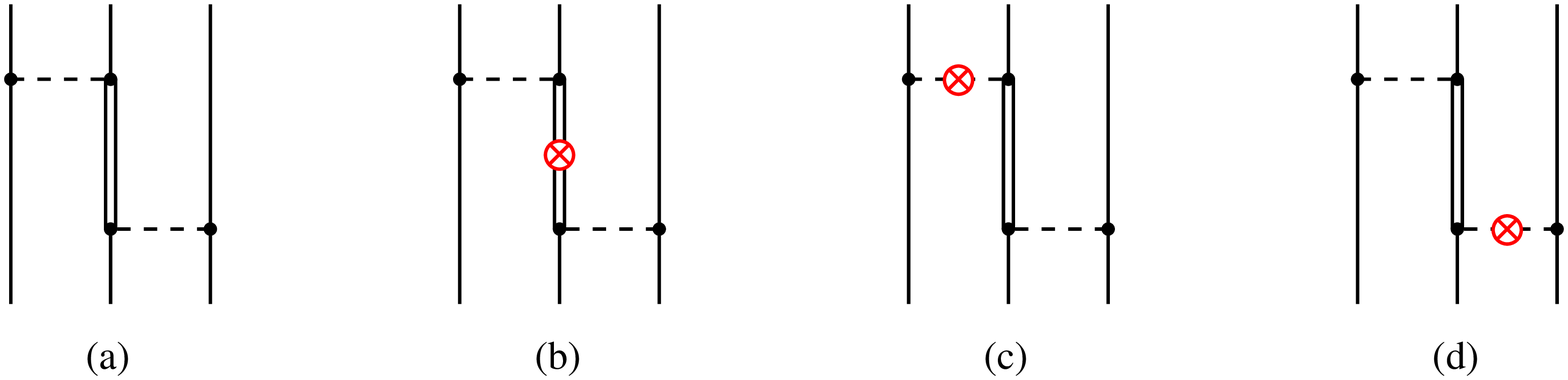}
    \caption{
         Leading isospin-conserving (diagram (a)) and isospin-breaking (diagrams (b-d))
         contributions to the 3NF with a $\Delta$-excitation. Solid, dashed
         and double lines represent nucleons, pions and deltas, respectively.
         Solid dots denote the leading isospin-invariant vertices, crossed
         circles are the isospin-breaking vertices from the
         $SU(2)_V$-rotated  effective Lagrangian as explained in section \ref{sec:deltamN}. 
\label{fig1} 
 }
\end{figure}

We are now in the position to discuss the 3NF contributions due to
intermediate $\Delta$ excitations. Due to vanishing of the lowest-order
contact interaction, the complete effect due the
$\Delta$  is given by a single two-pion ($2\pi$) exchange diagram (a)  in
Fig.~\ref{fig1}. 
It is well known that the corresponding 3NF
contribution is exactly reproduced at NNLO in EFT without explicit $\Delta$-fields via
resonance saturation of the low-energy constants (LECs) $c_{3,4}$ 
accompanying the subleading $\pi \pi NN$ vertices \cite{Bernard:1996gq}: $c_3 = - 2 c_4 = -{4}
h_A^2/(9 \Delta )$.  Here, $h_A$ denotes the
leading $\Delta N \pi$ axial-vector coupling constant.

Let us now regard the NNLO contributions. First, we emphasize that there are
still no diagrams involving contact interactions at this order. In fact, since
$NNN\Delta$ contact interactions contain at least two derivatives, they start
to contribute to the 3NF only at N$^3$LO. Therefore, the only diagrams which
need to be considered are the ones which result from graph (a) in
Fig.~\ref{fig1} by substituting one of the leading $\pi N \Delta$ vertices by
the subleading one which is proportional to the combination of LECs
$(b_3+b_8)$, see Refs.~\cite{Fettes:2000bb,Krebs:2007rh}.\footnote{Due to
  parity  invariance, the subleading $\pi NN$
vertices contain two derivatives or one insertion of $M_\pi^2$ more than the
leading one and thus need not be considered at NNLO.} Since this vertex
involves a time derivative, the corresponding 3NF contribution is shifted to
higher orders due to the $1/m_N$-suppression. We, therefore, end up with the
conclusion that the $2\pi$-exchange 3NF from diagram (a) in Fig.~\ref{fig1} 
represents the only additional contribution which arises in EFT with
explicit $\Delta$ up to NNLO.  

Before ending this section, it is interesting to compare the strength of various
pieces in the $2\pi$-exchange 3NF in EFT with and without explicit
$\Delta$ degrees of freedom which has the form
\beq
\label{3nftpe}
V_{\rm 2 \pi}=\sum_{i \not= j \not= k} \frac{1}{2}\left(
  \frac{g_A}{2 F_\pi} \right)^2 \frac{ \vec \sigma_i \cdot \vec q_{i}
  \; 
\vec \sigma_j \cdot \vec q_j  }{[q_i^2 + M_\pi^2 ][
q_j^2 + M_\pi^2]}  F^{\alpha \beta}_{k} \tau_i^\alpha 
\tau_j^\beta \,,
\eeq
where  $\vec q_i \equiv \vec p_i \, ' - \vec p_i$; $\vec p_i$
($\vec p_i \, '$) is the initial (final) momentum of the nucleon $i$ and 
\beq
\label{abcd}
F^{\alpha \beta}_{k} = \delta^{\alpha \beta} \left[ 
- a + b \, \vec q_i \cdot \vec q_j - c \, (q_i^2 + q_j^2) \right]
- d \, \epsilon^{\alpha \beta \gamma} \tau_k^\gamma  \; \vec q_i \times \vec
q_j \cdot \vec \sigma_k \,.
\eeq
The coefficients $a$, $b$ and $d$ can be expressed in terms of various
LECs while $c = 0$, see Ref.~\cite{Friar:1998zt} for more details. In EFT without explicit
$\Delta$, the chiral expansion for these coefficients starts at NNLO (i.e. at
order $Q^3$) where  one has:
\beq
\label{coefNoDelta}
a^{(3)} = \frac{4 c_1 M_\pi^2}{F_\pi^2}\,, \quad \quad
b^{(3)} = \frac{2 c_3}{F_\pi^2}\,, \quad \quad
d^{(3)} = - \frac{c_4}{F_\pi^2}\,. 
\eeq
Here the superscripts refer to the chiral order. 
In EFT with explicit $\Delta$ fields, the dominant contributions arise already at NLO
\beq
a^{(2)} = 0\,, \quad \quad
b^{(2)} = - \frac{8 h_A^2}{9\Delta F_\pi^2}\,, \quad \quad
d^{(2)} = - \frac{2 h_A^2}{9\Delta F_\pi^2}\,, 
\eeq
with the corrections at order $Q^3$ being still given by
Eq.~(\ref{coefNoDelta}). 
In Table \ref{tab:abd}, we compare the values for the coefficients $a$, $b$ and $d$ in
the theory with and without explicit $\Delta$ based on our determination
\cite{Krebs:2007rh} of the LECs $c_i$ and $b_3 + b_8$.
\begin{table*}[t] 
\begin{center}
\begin{tabular}{|ccccc|ccccc|cccccccccc|ccccc|}
\hline 
   &&  &&&&& $Q^3$, no $\Delta$ &&&&&  $Q^2$ with $\Delta$, fit 1 &&&&&
   $Q^2$ with $\Delta$, fit 2  &&&&&  $Q^3$ with $\Delta$, fits 1,2   && \\
\hline  \hline  
&& $a$  &&&&& $-$0.70   &&&&& ~~0   &&&&& ~~0 &&&&& $-$0.70 &&  \\
&& $b$  &&&&& $-$2.34  &&&&& $-$1.70  &&&&& $-$1.03 &&&&& $-$2.18  &&  \\
&& $d$  &&&&& $-$0.89   &&&&& $-$0.42   &&&&& $-$0.26 &&&&& $-$0.83  && \\
\hline 
  \end{tabular}
\caption{Values of the coefficients $a$, $b$ and $d$ entering the $2\pi$-exchange
  3NF in Eq.~(\ref{abcd}) in units of $M_\pi^{-1}$ for $a$ and $M_\pi^{-3}$
  for $b$ and $d$.
\label{tab:abd}}
\end{center}
\end{table*}
We remind the reader that fits 1 and 2 are based on the different values for
the $\pi N \Delta$ coupling constant $h_A$ used as an input: the
SU(4)/large-N$_c$ value $h_A = 3 g_A/(2 \sqrt{2}) = 1.34$ in fit 1 versus
$h_A = 1.05$ \cite{Fettes:2000bb} in fit 2. Clearly, at NLO this uncertainty in the
value of $h_A$  directly transforms into the uncertainty in the coefficients $b$ and
$d$, see the third and fourth columns in Table \ref{tab:abd}. At NNLO, however, both fits 1 and 2
lead to very similar results for the S- and P-wave $\pi N$ threshold parameters
\cite{Krebs:2007rh} and, consequently, for the coefficients $b$ and $d$. A
similar observation is also made for the subleading $2\pi$-exchange NN
potential in Ref.~\cite{Krebs:2007rh}. Another interesting result is that 
both EFTs with and without $\Delta$ lead to very similar values for the
coefficients $a$, $b$ and $d$ at NNLO. This seems to contradict the conclusion of
Ref.~\cite{Pandharipande:2005sx} where a significant overestimation of the
$\Delta$-contribution to the coefficients $b$ and $d$ by about
25\%  was found  in EFT without explicit $\Delta$ fields.
As shown in Table \ref{tab:abd}, the coefficients $b$ and $d$ are indeed
overestimated in the $\Delta$-less theory, but only by about 7$\%$. 
There are several reasons for this difference. 
First, our results for $b$ and
$d$ involve contributions beyond the ones generated by the $\Delta$
excitation. Secondly, we also take into account the subleading $\Delta$-contribution 
governed by the $b_3 + b_8$-vertex from $\mathcal{L}^{(2)}_{\pi N \Delta}$ which is not
considered in  Ref.~\cite{Pandharipande:2005sx}. 
Switching off the $b_3 + b_8$-interaction, however, tend to further decrease the above
mentioned overestimation and can, therefore, not explain the
discrepancy. The most important difference between our work and the one of
Ref.~\cite{Pandharipande:2005sx} concerns the determination of the LECs $c_3$
and $c_4$. In \cite{Pandharipande:2005sx}, this was achieved via matching the
P-wave $\pi N$ threshold parameters for $j=3/2$, $a_{1+}^+$ and $a_{1+}^-$. 
This leads to values for $b$ and $d$
which are quoted in Eq.~(15) of that work. On the other hand, in  
\cite{Krebs:2007rh} we used not only the $j=3/2$ but also the $j=1/2$ P-wave parameters 
$a_{1-}^+$ and $a_{1-}^-$ as well as the S-wave coefficients  $a_{0+}^+$ and $b_{0+}^+$ 
to pin down  the LECs $c_i$ and $b_3 + b_8$. This turns out to be the main reason
for the observed difference.

\section{Leading isospin-breaking effects}
\def\theequation{\arabic{section}.\arabic{equation}}
\setcounter{equation}{0}
\label{sec:isospin}

Our next goal is to study the leading isospin-breaking contributions to the
3NF. This can be done following the lines of
Refs.~\cite{Epelbaum:2005fd,Epelbaum:2004xf} where we worked out the leading and subleading 
isospin-breaking 2N and 3N potentials, see also 
Refs.~\cite{Friar:1999zr,Friar:2003yv,Friar:2004rg,Friar:2004ca} for a related
work on this subject. Here and in what follows, we adopt the same counting
rules for the electric charge $e$ and the quark mass ratio
$\epsilon = ( m_u-m_d )/ (m_u+m_d ) \sim -1/3$  as in 
Refs.~\cite{Epelbaum:2005fd,Epelbaum:2004xf}, namely
\beq\label{CountRules}
\epsilon \sim e \sim \frac{Q}{\Lambda}; \quad \quad
\frac{e^2}{(4 \pi )^2}  \sim \frac{Q^4}{\Lambda^4}\,.
\eeq
The leading isospin-breaking vertices have the dimension $\Delta_i = 2$ and
correspond to the charged-to-neutral pion and proton-to-neutron mass
differences \cite{Epelbaum:2005fd} and, as will be shown below, mass splittings between the 
different charge states of the $\Delta$. It is easy to verify using 
Eqs.~(\ref{powc}) and (\ref{chirdim}) that the dominant
isospin-breaking 3NF which involves an intermediate delta excitation results at
order $\nu = 4$ from isospin-breaking pion-, nucleon- and delta-mass
insertions in the first diagram in Fig.~\ref{fig1}. Notice that the leading
isospin-breaking $\pi \pi NN$ vertex of dimension $\Delta_i = 2$ also generates the
3NF contribution at order $\nu =4$ \cite{Epelbaum:2004xf,Friar:2004rg} which,
however, does not involve
$\Delta$-excitations and is, therefore, irrelevant for the present study.  
Thus, in order to proceed with the calculation of the leading isospin-breaking 3NF
contributions due to intermediate $\Delta$-excitations, we first need to
analyze the delta mass
splittings in chiral effective field theory.

\subsection{Delta mass splittings in chiral effective field theory}
\label{sec:deltamN}

To analyze the delta mass splitting in chiral effective field theory, we 
need to include  hard virtual photons in the effective pion-nucleon-delta Lagrangian. 
Specifically, we are interested in the
leading-order virtual photon effects that show up as local operators
of dimension $\Delta_i = 3$. To construct these terms, we employ the standard
spurion method, 
see e.g. \cite{Urech:1994hd,Meissner:1997ii,Muller:1999ww,Gasser:2002am}.
The resulting Lagrangian reads
\beqa\label{L2em}
{\cal L}_{\Delta \gamma}^{\Delta_i=3} &=& - \bar{T}_i^\mu \, F_\pi^2\, \biggl[ 
  f_1^\Delta \, \delta_{ij} \, \langle Q_+^2 - Q_-^2\rangle
+ f_2^\Delta \, \delta_{ij} \, \langle Q_+ \rangle Q_+
+ f_3^\Delta \, \delta_{ij} \, \langle Q_+ \rangle^2 \nonumber \\
&& \qquad\quad + f_4^\Delta \, \langle \tau^i Q_+ \rangle\langle \tau^j Q_+ \rangle
+ f_5^\Delta \, \langle \tau^i Q_- \rangle\langle \tau^j Q_- \rangle
\, \biggr] \, g_{\mu\nu} \, T^\nu_j
\eeqa
where $\langle \ldots \rangle$ denotes the isospin trace,
the $Q_\pm$ are defined as in \cite{Meissner:1997ii} and $F_\pi$ is the pion 
decay constant in the chiral limit. The factor $F_\pi^2$ ensures
that the LECs $f_i^\Delta$ have the same dimension as the corresponding strong
LECs \cite{Meissner:1997ii}. 
Notice that while all couplings and masses appearing in the effective
Lagrangian should, strictly speaking, be
taken at their chiral limit values, to the accuracy we are working, 
we can use their pertinent physical values. We further emphasize that the
expected natural size of the LECs $f_i^\Delta$ is 
\beq
\label{natural_units}
F_\pi^2 \, f_i^\Delta \sim
M_\rho \, \frac{1}{(4 \pi)^2},
\eeq
so that, according  to Eqs.~(\ref{chirdim}) and (\ref{CountRules}), the
corresponding vertices are indeed of  the dimension $\Delta_i =3$. 
The electromagnetic mass term
of the delta is readily deduced from Eq.~(\ref{L2em}) by considering the
terms without pion fields, e.g.
\beqa\label{massem}
{\cal L}_{\Delta \gamma, \, {\rm mass}}^{\Delta_i=3} &=& - F_\pi^2 \, \bar{T}_i^\mu \, \biggl[
(f_1^\Delta + f_3^\Delta) \, e^2 \,  \delta_{ij} + f_2^\Delta \, \frac{e^2}{2}
(1+\tau^3)  \,  \delta_{ij} + f_4^\Delta \, e^2 \, \delta_{i3}\, \delta_{j3}
\, \biggr] \, g_{\mu\nu} \, T^\nu_j ~.
\eeqa
Note that the first term in this equation leads to an overall mass shift in
the delta quartet, while the other two contribute to various splittings.
The relevant isospin conserving and strong isospin violating terms at leading
order are given by \cite{Hemmert:1997ye}
\beq\label{massstr}
{\cal L}_{\Delta \, {\rm IV}}^{\Delta_i=2} =  - \bar{T}_i^\mu \,
c_5^\Delta \, \left(\chi_+ - \langle \chi_+\rangle \right) \,  \delta_{ij} \,
 g_{\mu\nu} \, T^\nu_j =  - \bar{T}_i^\mu \, c_5^\Delta \,2M_\pi^2 \, \varepsilon \, \tau^3 \,
 \delta_{ij} \, g_{\mu\nu} \, T^\nu_j  + \ldots ~
\eeq
where ellipses refer to terms which contain pion fields. 
Note that in 
Ref.~\cite{Hemmert:1997ye} the LEC $c_5^\Delta$ was denoted as $a_5$.
Combining Eqs.~(\ref{massem}) and (\ref{massstr}), we arrive at the leading
strong and electromagnetic isospin-breaking contributions to the delta mass
term
\beq
\label{deltamass}
{\cal L}_{\Delta, \, {\rm mass,} \,  {\rm IV}} = - \bar{T}_i^\mu \, \biggl[
\, - \delta m_{\Delta}^1\frac{1}{2}\, \tau^3
\,  \delta_{ij} - \delta m_{\Delta}^2\frac{3}{4} \,  \delta_{i3} \, \delta_{j3}
\, \biggr] \, g_{\mu\nu} \, T^\nu_j~, 
\eeq
with  
\beqa
\label{deltas}
\delta m_\Delta^1&=&-4 M_\pi^2 \, \epsilon \, c_5^\Delta -   F_\pi^2 \, e^2 \,
f_2^\Delta~, \nonumber \\
\delta m_\Delta^2 &=& -\frac{4}{3} F_\pi^2 \, e^2 \, f_4^\Delta ~.
\eeqa
Notice that according to the counting rules in Eq.~(\ref{CountRules}), the
dominant term in Eq.~(\ref{deltamass}) arises from strong isospin breaking while the
leading electromagnetic contribution is shifted one order
higher. At least in the nucleon sector, this pattern is in a reasonable
agreement with the data and the strong shift gives indeed numerically the dominant
contribution. Since we are interested here only in the leading
isospin-breaking effects due to explicit $\Delta$ degrees of freedom, it is, strictly speaking,
sufficient to keep only the strong delta mass shift. We will, however,
keep both the leading strong and electromagnetic delta mass shifts in what
follows and thus generate a portion of the subleading isospin-breaking
contributions to the 3NF. 
At the order we are working, isospin-breaking contribution to the delta
self-energy results from the tree diagram with a single insertion from
Eq.~(\ref{deltamass}). The delta mass splittings can, therefore, be directly
read off from Eq.~(\ref{deltamass}). Switching from the Rarita-Schwinger
representation to
physical delta fields $( \Delta^{++}, \,  \Delta^{+}, \,   \Delta^{0}, \,
\Delta^{-})$ one obtains:
\beqa
m_{\Delta^{++}} &=& \tilde{m}_{\Delta} +\frac{1}{2}\,\delta m_\Delta^1~,  \nonumber \\
m_{\Delta^{+}}  &=& \tilde{m}_{\Delta} + \frac{1}{6}\,\delta
m_\Delta^1 + \frac{1}{2} \delta m_\Delta^2~, \nonumber \\
m_{\Delta^{0}}  &=&  \tilde{m}_{\Delta}  - \frac{1}{6}\,\delta
m_\Delta^1 + \frac{1}{2} \delta m_\Delta^2~, \nonumber \\
m_{\Delta^{-}}  &=& \tilde{m}_{\Delta} - \frac{1}{2}\,\delta m_\Delta^1~. 
\eeqa
Notice that the isospin-invariant delta mass shift   $\delta m_\Delta$,
defined according to $ \tilde{m}_{\Delta} = \krig{m}_\Delta + \delta
m_\Delta$, can, in principle, be extracted using  the chiral limit of the 
$N\Delta$ splitting $\Delta_0 \simeq
330\,$MeV from Ref.~\cite{Bernard:2005fy} and the value of the nucleon mass
in the chiral limit $\krig{m}_N \simeq 890\,$MeV from Ref.~\cite{Bernard:2003rp}.
This leads to $\krig m_\Delta \simeq 1220\,$MeV. 
We further emphasize that in the absence of electromagnetic corrections 
(no  $\delta m_\Delta^2$-term), there is an equal spacing between the members
of the quartet. In that case,
one has $\bar{m}_{\Delta} = \tilde{m}_{\Delta}$ and all splittings are equal
to $\delta m_\Delta^1/3$. 
We also recover the SU(6) results $m_{\Delta^{++}} -
m_{\Delta^{-}} = 3( m_{\Delta^+} - m_{\Delta^0})$ independent of the strength
of the electromagnetic LEC $f_\Delta^4$, see e.g. \cite{Rubinstein:1967aa}.
Our results for the strong splittings are, of course, in agreement with the
ones of Ref.\cite{Tiburzi:2005na}. In that paper, higher order corrections
were also evaluated with particular emphasis on the quark mass dependence of
the delta splittings to be used as chiral extrapolation functions in lattice
gauge theory.

To further analyze the delta mass splittings, we need input values for some
of these masses. Astonishingly, the available information of these masses
is fairly scarce and uncertain, see e.g. the most recent listings of the
particle data group \cite{PDG}.
{}From pion-nucleon 
scattering, one can extract $m_{\Delta^{++}}$ and $m_{\Delta^{0}}$ as well as
the average delta mass $\bar m_\Delta$ 
\beq
\label{average_mass}
\bar{m}_{\Delta} \equiv \frac{1}{4} \, \left( m_{\Delta^{++}} +  m_{\Delta^{+}} +
m_{\Delta^{0}} +  m_{\Delta^{-}} \right) =  \tilde{m}_{\Delta} + \frac{1}{4}\delta m_\Delta^2~.
\eeq
The values for the  Breit-Wigner masses $m_{\Delta^{++}}$ and $m_{\Delta^{0}}$ quoted in
Ref.~\cite{PDG} and based on the determinations from
Refs.~\cite{Gridnev:2004mk,Abaev:1995cx,Koch:1980ay,Pedroni:1978it} are all in
a reasonable
agreement with each other (within the given error bars). The only exception
is the analysis of Ref.~\cite{Bernicha:1995gg}, which yields
significantly different  values
for $m_{\Delta^{++}}$ and $m_{\Delta^{0}}$ (but has also the largest
error bars).  From photoproduction reactions, one
can, in principle, determine the mass of $\Delta^+$. The two available
results for $\Delta^+$ quoted in \cite{PDG} differ, however, significantly
from each other. We, therefore, refrain from using $m_{\Delta^{+}}$ in our
study. Finally, no experimental information is available for
$m_{\Delta^{-}}$. To  pin down the values for $\tilde m_\Delta$,
$\delta m_\Delta^1$ and $\delta m_\Delta^2$ we proceed in two different ways. 
First, we use as an input the available data on $m_{\Delta^{++}}$, 
$m_{\Delta^{0}}$ (except the values from
Ref.~\cite{Bernicha:1995gg}) and the average delta mass $\bar{m}_{\Delta}$.  
For the latter, we adopt the value $\bar{m}_{\Delta}=1233$ MeV which is
consistent with the estimation of Ref.~\cite{PDG},  $\bar{m}_{\Delta}
= 1231 \ldots 1233$ MeV as well as with the most recent determination
from Ref.~\cite{Arndt:2006bf}, $\bar m_\Delta = 1233.4 \pm 0.4$ MeV.  
This leads to 
\beq
\label{fit1}
\tilde{m}_{\Delta} = 1233.4 \pm 0.7\mbox{ MeV}\,, \quad \quad
\delta m_\Delta^1 = -5.3 \pm 2.0\mbox{ MeV}\,, \quad \quad
\delta m_\Delta^2 = -1.7 \pm 2.7\mbox{ MeV}\,,
\eeq
where the error bars result from using different input values for 
$m_{\Delta^{++}}$ and $m_{\Delta^{0}}$, see Table \ref{tab2a}.  
\begin{table*}[h] 
\begin{center}
\begin{tabular}{|ccccccc|ccccccccc|ccc|} 
\hline 
&$\tilde{m}_{\Delta}$ &&  $\delta m_\Delta^1$ &&  $\delta m_\Delta^2$
&&& $m_{\Delta^{++}}$  && $m_{\Delta^{+}}$ 
&& $m_{\Delta^{0}}$ &&  $m_{\Delta^{-}}$ 
&&& input &\\
\hline \hline 
&1233.63 &&  -6.15  &&  -2.50 &&& 1230.55$^*$ && 1231.35 && 1233.40$^*$ && 1236.70
&&& \protect\cite{Gridnev:2004mk} &\\
&1234.10 &&  -7.20  &&  -4.40 &&& 1230.50$^*$ && 1230.70 && 1233.10$^*$ && 1237.70 
&&& \protect\cite{Abaev:1995cx} &\\
&1233.15 &&  -4.50  &&  -0.60 &&& 1230.90$^*$ && 1232.10 && 1233.60$^*$ && 1235.40  
&&& \protect\cite{Koch:1980ay} &  \\
&1232.75 &&  -3.30  &&  1.00 &&& 1231.10$^*$ && 1232.70 && 1233.80$^*$ && 1234.40  
&&& \protect\cite{Pedroni:1978it} &  \\
\hline 
   \end{tabular}
\caption{
Delta masses and LECs for various input values of $m_{\Delta^{++}}$ and
$m_{\Delta^{0}}$  as indicated by the star. Additional input is
the average delta mass $\bar{m}_{\Delta}=1233$ MeV. All values are given in units
of MeV. 
\label{tab2a}}
\end{center} 
\end{table*}
Notice that the obtained results for $\delta m_\Delta^1$ and $\delta
m_\Delta^2$ are of a natural size. Indeed, based on the naive dimensional
analysis, one expects 
$| \delta  m_\Delta^1 | \sim | \epsilon M_\pi^2 /M_\rho | \simeq 8$ MeV  and 
$| \delta  m_\Delta^2 | \sim e^2 M_\rho /(4 \pi)^2  \simeq 1.5$ MeV.  

As an alternative, one can use in the determination of 
$\tilde m_\Delta$, $\delta m_\Delta^1$ and $\delta m_\Delta^2$
the quark model relation  \cite{Rubinstein:1967aa}
\beq
\label{furtherinput}
m_{\Delta^+} - m_{\Delta^0} = m_p - m_n 
\eeq
instead of the average delta mass  $\bar{m}_{\Delta}$. 
This fixes the value of $\delta m_\Delta^1$ and leads to less
uncertain results for $\tilde m_\Delta$ and $\delta m_\Delta^2$, see Table \ref{tab2}
for more details:
\begin{table*}[h] 
\begin{center}
\begin{tabular}{|ccccc|ccccccccc|ccc|ccc|} 
\hline 
&$\tilde{m}_{\Delta}$ &&  $\delta m_\Delta^2$
&&& $m_{\Delta^{++}}$  && $m_{\Delta^{+}}$ 
&& $m_{\Delta^{0}}$ &&  $m_{\Delta^{-}}$  &&& $\bar{m}_{\Delta}$
&&& input &\\
\hline \hline 
&1232.49 &&  0.53 &&& 1230.55$^*$ && 1232.11 && 1233.40$^*$ && 1234.43
&&& 1232.62 &&& \protect\cite{Gridnev:2004mk} &\\
&1232.44 &&  0.03 &&& 1230.50$^*$ && 1231.81 && 1233.10$^*$ && 1234.38 
&&& 1232.45 &&& \protect\cite{Abaev:1995cx} &\\
&1232.84 &&  0.23 &&& 1230.90$^*$ && 1232.31 && 1233.60$^*$ && 1234.78  
&&& 1232.90 &&& \protect\cite{Koch:1980ay} &  \\
&1233.04 &&  0.23 &&& 1231.10$^*$ && 1232.51 && 1233.80$^*$ && 1234.98  
&&& 1233.10 &&& \protect\cite{Pedroni:1978it} &  \\
\hline 
   \end{tabular}
\caption{
Delta masses and LECs for various input values of $m_{\Delta^{++}}$ and
$m_{\Delta^{0}}$ as indicated by the star. Additional input is
the quark model relation (\ref{furtherinput}). All values are given in units
of MeV. 
\label{tab2}}
\end{center} 
\end{table*}
\beq
\tilde{m}_{\Delta} = 1232.7 \pm 0.3\mbox{ MeV}\,, \quad \quad
\delta m_\Delta^1 = -3.9 \mbox{ MeV}\,, \quad \quad
\delta m_\Delta^2 = 0.3 \pm 0.3\mbox{ MeV}\,.
\eeq
It is comforting to see that the results of both determinations are compatible
with each other. In particular, the value obtained for the average delta mass, $\bar
m_\Delta = 1232.45 
\ldots 1233.10$ MeV, agrees with both the estimation of Ref.~\cite{PDG} 
and the determination of Ref.~\cite{Arndt:2006bf}. We further emphasize 
that the LECs
$c_5^\Delta$ and $f_2^\Delta$ can be deduced from $\delta m_N^{\rm str}$ and
$\delta m_N^{\rm em}$ if one uses the relation
(\ref{furtherinput}) separately for the strong and electromagnetic mass
shifts. Using $\delta m_N^{\rm str} \equiv (m_p - m_n)^{\rm str} = -2.05 \pm
0.3$ MeV and   
$\delta m_N^{\rm em} \equiv (m_p - m_n)^{\rm em} = 0.76 \pm
0.3$ MeV from Ref.~\cite{Gasser:1982ap}, see also \cite{Beane:2006fk} for a recent
determination from lattice QCD, one obtains 
$c_5^\Delta = 3 c_5 = - 0.24 \pm 0.04$ GeV$^{-1}$  and 
$f_2^\Delta = 3 f_2 = - 2.9 \pm 1.1$ GeV$^{-1}$, where $c_5$ and $f_2$ are the
corresponding LECs in the nucleon sector.

\subsection{Field redefinitions and the proton-to-neutron mass shift}
\def\theequation{\arabic{section}.\arabic{equation}}
\label{sec:Nmass}

Having worked out the delta mass splitting in chiral EFT, it is, in principle,
a straightforward task to calculate the dominant isospin-breaking 3NF contribution
due to explicit $\Delta$ following the line of Ref.~\cite{Epelbaum:2005fd}. 
The calculations may be facilitated if one eliminates the
nucleon mass shift from the effective Lagrangian. This allows to get rid of
diagrams which involve reducible topologies and, therefore, enables to use the
Feynman graph technique to derive the 3NF. As already demonstrated  in
Ref.~\cite{Friar:2004ca} for EFT without explicit $\Delta$, the
proton-to-neutron  mass
difference can be eliminated  from $\mathcal{L}_{\rm eff}$ via a suitable
redefinition of the pion and nucleon fields (or, equivalently, via a suitable 
\emph{local} $SU(2)_{\rm V}$ transformation) in favor of new
vertices proportional to $\delta m_N$. This approach can be
straightforwardly generalized to include the $\Delta$ as an explicit degree of
freedom. To be specific, consider the effective chiral Lagrangian 
$\mathcal{L}_{\rm eff} ( \Phi , \, J )$,  where 
$\Phi \equiv \{  U,  \bar N,  N, \bar T,  T \}$
($J \equiv \{  r_\mu,  l_\mu,  s, p \}$) collectively denote the pion, nucleon and
delta fields (right- and left-handed, scalar and pseudoscalar external
sources). The effective Lagrangian  $\mathcal{L}_{\rm eff} ( \Phi , \, J )$ is
invariant under local chiral rotations $G = SU(2)_L \times SU(2)_R$:
\beq
\mathcal{L}_{\rm eff} (\Phi , \, J)  \stackrel{g \in G }{\longrightarrow}
\mathcal{L}_{\rm eff} (\Phi ', \, J ' ) \equiv  \mathcal{L}_{\rm eff} (g (\Phi
), \, g (J)  ) = \mathcal{L}_{\rm eff} (\Phi , \, J ) \,.
\eeq
For the purpose of computing S-matrix elements in the few-nucleon sector, the
external sources can be set to zero from the beginning (with the exception of
the scalar source that is equal to the quark mass matrix in this limit). 
In the absence of external sources, the effective Lagrangian $\mathcal{L}_{\rm
  eff} (\Phi  ) \equiv  \mathcal{L}_{\rm eff} (\Phi , \, J  ) \big|_{J=0}$
is only invariant under global chiral rotations. Local
chiral transformations $g \in G$ will, in  general, affect its form
\beq
\mathcal{L}_{\rm eff} (\Phi )  \stackrel{g \in G }{\longrightarrow}
\mathcal{L}_{\rm eff} (g ( \Phi ) ) =  \mathcal{L}_{\rm eff} ( \Phi  ) 
+ \delta \mathcal{L}_{\rm eff} ( \Phi  )\,,
\eeq 
and obviously may be viewed as a redefinition of fields $\Phi$. The task is
now to choose the
transformation $V$ in such a way that the resulting correction $\delta \mathcal{L}_{\rm eff}
( \Phi )$ eliminates the nucleon mass shift term  
\beq
- \bar{N}\delta m_N \frac{1}{2}\tau^3 N,\quad \quad 
\delta m_N=-4\, c_5 \,\epsilon\, M_\pi^2 - e^2 F_\pi^2 f_2\, ,
\label{PropCorrIsoBr}
\eeq
in $\mathcal{L}_{\rm eff} ( \Phi )$. Notice that in practice, it is more
convenient to compute the correction $\delta
\mathcal{L}_{\rm eff}$ applying the inverse rotation to the external
currents 
\beq
\mathcal{L}_{\rm eff} (g ( \Phi ) ) = \mathcal{L}_{\rm eff} \left( \Phi , \, g^{-1}
(J)\right)\Big|_{J = 0}  =  \mathcal{L}_{\rm eff} ( \Phi  ) +  \delta \mathcal{L}_{\rm eff} ( \Phi  )\,.
\eeq
Following Ref.~\cite{Friar:2004ca}, we choose the local $SU(2)_V$
transformation such that 
\beq
r_\mu^\prime = V r_\mu V^\dagger + i V\partial_\mu V^\dagger,\quad \quad 
l_\mu^\prime = V l_\mu V^\dagger + i V\partial_\mu V^\dagger,\quad \quad
s^\prime=V s V^\dagger, \quad \quad 
p^\prime = V p V^\dagger
\eeq
with $V=\exp(i\,v\cdot x\,\delta m_N\tau^3/2)$.  
Notice that as pointed out in Ref.~\cite{Friar:2004ca}, this transformation
does not lead to explicitly $x$-dependent vertices  since the
interactions in $\mathcal{L}_{\rm eff}$ conserve electric charge.   
One obtains for the chiral vielbein $u_\mu = 
i \left[ u^\dagger (\partial_\mu - i r_\mu ) u 
- u (\partial_\mu - i l_\mu ) u^\dagger \right] $
and connection $\Gamma_\mu =1/2  \left[ u^\dagger (\partial_\mu - i r_\mu ) u 
+ u (\partial_\mu - i l_\mu ) u^\dagger \right] $
\beqa
u_\mu \Big|_{J=0} \to i \left( u^\dagger (\partial_\mu - i r_\mu ') u 
- u (\partial_\mu - i l_\mu ') u^\dagger \right) \Big|_{J=0} &=& 
u_\mu \Big|_{J=0} + \frac{\delta m_N}{2} \,  v_\mu \left( u^\dagger \tau^3
  u -  u \tau^3  u^\dagger \right) \nn
&=& u_\mu \Big|_{J=0} + \frac{\delta m_N}{F_\pi} v_\mu [\fet \tau \times \fet \pi
]_3 + \mathcal{O} ( \fet{\pi}^3 )
\,, \nn
\Gamma_\mu \Big|_{J=0} \to \frac{1}{2} \left( u^\dagger (\partial_\mu - i r_\mu ') u 
+ u (\partial_\mu - i l_\mu ') u^\dagger \right) \Big|_{J=0} &=& 
\Gamma_\mu \Big|_{J=0} -  i \frac{\delta m_N}{4} \,  v_\mu \left( u^\dagger \tau^3
  u +  u \tau^3  u^\dagger \right) \\
&=& 
\Gamma_\mu \Big|_{J=0} -  i \frac{\delta m_N}{2} v_\mu \tau^3  
+ i \frac{\delta m_N}{4F_\pi^2} v_\mu [\fet \pi^2 \tau^3 - \fet \pi \cdot \fet \tau
\pi_3 ]  + \mathcal{O} ( \fet \pi^4 )\,,
\nonumber
\eeqa
and, consequently, for the nucleon and delta kinetic energy terms:
\beqa
\bar N i v \cdot D N &=& \bar N \left( i v \cdot \partial + i v \cdot \Gamma
\right) N \to \bar N \left( 
i v \cdot \partial + \frac{\delta m_N}{2} \tau^3 + \mathcal{O} (\fet \pi^2)
 \right) N \,, \nn
- \bar T_i^\mu \left( i v \cdot D_{ij} \right) g_{\mu \nu } T_j^\nu 
&=& - \bar T_i^\mu \big( i (v \cdot \partial + v \cdot \Gamma ) \, \delta_{ij} 
+ \epsilon_{ijk} \langle \tau^k v \cdot \Gamma \rangle 
\big) \, g_{\mu \nu } T_j^\nu \nn
&\to& 
- \bar T_i^\mu \left( \left( i v \cdot \partial 
+ \frac{3 \delta m_N}{2} \tau^3 \right) \delta_{ij} + \mathcal{O} (\fet \pi^2)
\right) \, g_{\mu \nu } T_j^\nu \,. 
\eeqa
Here we have used the following identity for the Rarita-Schwinger field:
\beq
\bar T^\mu_i \epsilon_{ijk} T^\nu_j = i \bar T^\mu_i \delta_{ij} \tau^k T^\nu_j\,.
\eeq
Thus, we finally end up with the modification of the lowest-order effective Lagrangian
\beqa
\mathcal{L}_{\rm eff}^{\Delta_i = 0} &\to& 
\mathcal{L}_{\rm eff}^{\Delta_i = 0} - \delta m_N [ (v \cdot \partial \fet \pi
) \times \fet \pi ]_3 + \frac{1}{2} (\delta m_N)^2 (\fet \pi^2 - \pi_3 \pi_3 )
+ \mathcal{O}(\fet \pi^4)
+ \frac{\delta m_N}{2} \bar N \left( \tau^3 + \mathcal{O}(\fet \pi^2) \right) N \nn
&& \mbox{\hskip 0.95 true cm} - \frac{3 \delta m_N}{2} \bar T_i^\mu \left( \tau^3 \, \delta_{ij} +
  \mathcal{O}(\fet \pi^2) \right) \, g_{\mu \nu} T_j^\mu\,. 
\eeqa
For the pion and pion-nucleon sectors, these results agree with the ones
obtained in Ref.~\cite{Friar:2004ca}. 

Finally, we would like to emphasize that
although the applied field redefinition is, strictly speaking, of a more general
type than the one discussed in \cite{Haag:1958vt,Coleman:1969sm}, it is straightforward to show that
both $\mathcal{L}_{\rm eff}$ and $\mathcal{L}_{\rm eff} + \delta \mathcal{L}_{\rm
  eff}$ lead to the same $S$-matrix elements in the few-nucleon sector. 
The corresponding Green functions in the original and modified theories are,
however, not the  same\footnote{For
  example, the free nucleon propagator in the modified theory does not contain
  $\delta m_N$  any more.} and related to
each other by the transformations $V$: 
\beqa
\frac{\delta^{2n}Z[J,\eta,\bar{\eta}]}{\delta
  \eta_{j_1}(x_1)\dots\delta \eta_{j_n}(x_n)\delta\bar{\eta}_{i_1}(y_1)\dots
\delta\bar{\eta}_{i_n}(y_n)}\bigg|_{J=\eta=\bar{\eta}=0}&=&
\sum_{k_1,l_1,\dots,k_n,l_n}
\frac{\delta^{2n}Z[J^\prime,\eta^\prime,
\bar{\eta}^\prime]}{\delta
  \eta_{k_1}^\prime(x_1)\dots\delta \eta_{k_n}^\prime(x_n)\delta\bar{\eta}_{l_1}^\prime(y_1)\dots
\delta\bar{\eta}_{l_n}^\prime(y_n)}\bigg|_{J^\prime=\eta^\prime=\bar{\eta}^\prime=0}\nn
&\times&V_{k_1,j_1}(x_1)\dots V_{k_n,j_n}(x_n)V_{i_1,l_1}^\dagger(y_1)
\dots V_{i_n,l_n}^\dagger(y_n)\label{GreenPrime} \,.
\eeqa
Here $Z[J,\eta,\bar{\eta}]$ is the generating functional 
\beq
e^{i Z[J,\eta,\bar{\eta}]}=
\int [D \Phi ]\, e^{i\int d^4 x [\mathcal{L}_{\rm eff}(\Phi, J)
+\bar{\eta} N+\bar{N} \eta ]} \,,
\eeq
and $\eta^\prime = V \eta$, $\bar{\eta}^\prime=\bar{\eta}V^\dagger$.

\subsection{Isospin-breaking 3NFs due to explicit deltas}
\label{sec:3NFisov}

We are now in the position to present our results for the leading
isospin-breaking contributions to the 3NF due to an intermediate
$\Delta$-excitation. A straightforward evaluation of Feynman diagrams (b-d)
in Fig.~\ref{fig1} yields the following charge-symmetry conserving (CSC),
i.e.~class-II in the 
notation of Ref.~\cite{Epelbaum:2004xf}, and and charge-symmetry breaking (CSB),
i.e.~class-III, contributions to the 3NF:
\beqa
\label{res_mom}
V_{\rm 3N}^{\rm Class-II}&=& - \sum_{i\neq j\neq k} \frac{g_A^2 h_A^2}{18 F_\pi^4
  \Delta} \, \frac{\vec{q}_i\cdot \vec{\sigma}_i \vec{q}_k\cdot
\vec{\sigma}_k}{[q_i^2 + M_\pi^2][q_k^2 + M_\pi^2]} \bigg( \tau_i^3 \tau_k^3 \;
\bigg( \frac{4 \delta M_\pi^2 }{q_k^2 + M_\pi^2} - \frac{3 \delta m_\Delta^2}{4 \Delta}
\bigg) \, \vec q_i \cdot \vec q_k \nn
&& {} 
+ \bigg( \frac{\delta M_\pi^2 }{q_k^2 + M_\pi^2} \; 
[\fet \tau_i \times \fet \tau_j ]^3 \tau_k^3  - \frac{3 \delta m_\Delta^2}{8 \Delta}
[\fet \tau_i \times \fet \tau_k ]^3 \tau_j^3  \;
\bigg) \, [\vec q_i \times \vec q_k ] \cdot \vec \sigma_j  \bigg)\,,\nn [3pt]
V_{\rm 3N}^{\rm Class-III}&=& - \sum_{i\neq j\neq k} \frac{g_A^2 h_A^2
  (\delta m_\Delta^1 - 3 \delta m_N )}{216 F_\pi^4
  \Delta^2} \, \frac{\vec{q}_i\cdot \vec{\sigma}_i \vec{q}_k\cdot
\vec{\sigma}_k}{[q_i^2 + M_\pi^2][q_k^2 + M_\pi^2]} \Big( \, 5 \,[ \fet \tau_i \times
\fet \tau_k ]^3 \; [\vec q_i \times \vec q_k ] \cdot \vec \sigma_j \nn
&& {} + 4 (\fet{\tau}_j\cdot \fet{\tau}_k \tau_i^3-2 \fet{\tau}_i\cdot \fet{\tau}_k
\tau_j^3 ) \;\vec{q}_i\cdot
    \vec{q}_k \Big)\,.
\eeqa
The above expressions together with  Eqs.~(46), (49), (52) and (54) of
Ref.~\cite{Epelbaum:2004xf} corresponding to the contributions 
from  $2\pi$-exchange diagrams without intermediate delta-excitation provide
the leading isospin-breaking 3NF in EFT with explicit $\Delta$. 
 
The obtained results can be straightforwardly transformed into configuration
space: 
\beqa
\label{r_space}
V_{\rm 3N}^{\rm Class-II}&=&  - \sum_{i\neq j\neq k} \frac{g_A^2 h_A^2 M_\pi^6}{288
  \pi^2 F_\pi^4 \Delta} \, \vec \sigma_i \cdot \vec \nabla_{ij} \; \vec \sigma_k \cdot \vec \nabla_{kj}
\bigg[ \tau_i^3 \tau_k^3 \; \vec \nabla_{ij} \cdot \vec \nabla_{kj} \; 
\bigg( \frac{2 \delta M_\pi^2}{M_\pi^2} \, U_1 ( x_{ij}) \, U_2 ( x_{kj}) - \frac{3 \delta m_\Delta^2}{4 \Delta}
\, U_1 ( x_{ij}) \, U_1 (x_{kj}) \bigg) \nn
&& {}+ 
[ \vec \nabla_{ij} \times \vec \nabla_{kj} ] \cdot \vec \sigma_j
\bigg(  \frac{\delta M_\pi^2 }{2 M_\pi^2} \, [\fet \tau_i \times \fet \tau_j ]^3 \tau_k^3   \, U_1 ( x_{ij}) \, U_2 ( x_{kj}) 
- \frac{3 \delta m_\Delta^2}{8 \Delta} \, [\fet \tau_i \times \fet \tau_k ]^3 \tau_j^3  
\, U_1 ( x_{ij}) \, U_1 (x_{kj}) 
\bigg) \bigg]\,, \nn
V_{\rm 3N}^{\rm Class-III}&=&  - \sum_{i\neq j\neq k} \frac{g_A^2 h_A^2 
  (\delta m_\Delta^1 - 3 \delta m_N ) M_\pi^6}{3456 \pi^2 F_\pi^4
  \Delta^2} \,  \vec \sigma_i \cdot \vec \nabla_{ij} \; \vec \sigma_k \cdot \vec \nabla_{kj}
\Big[ 5 [ \fet \tau_i  \times \fet \tau_k ]^3 \; [ \vec \nabla_{ij} \times
\vec \nabla_{kj} ] \cdot \vec \sigma_j \nn
&& {} + 4 (\fet{\tau}_j\cdot \fet{\tau}_k \tau_i^3-2 \fet{\tau}_i\cdot \fet{\tau}_k
\tau_j^3 ) \; \vec \nabla_{ij} \cdot \vec \nabla_{kj}  \; \Big] \,  U_1 ( x_{ij}) \, U_1 (x_{kj}) \,,
\eeqa
where $\vec r_{ij} \equiv \vec r_i - \vec r_j$ is the distance between the
nucleons $i$ and $j$,  $\vec x_i \equiv M_\pi \, \vec r_i$, $\vec \nabla_i$ act on $\vec x_i$ and 
$x_{ij} \equiv | \vec x_{ij}  |$. Further, we have introduced the profile
functions  
\beqa
U_1 (x) &=& \frac{4 \pi }{M_\pi} \int \frac{d^3 q}{(2 \pi )^3} \, \frac{e^{i
    \vec q \cdot \vec x/M_\pi}}{q^2 + M_\pi^2} = \frac{e^{-x}}{x}\,, \nn
U_2 (x) &=& 8 \pi M_\pi \,  \int \frac{d^3 q}{(2 \pi )^3} \, \frac{e^{i
    \vec q \cdot \vec x/M_\pi}}{[q^2 + M_\pi^2]^2} = e^{-x}\,.
\eeqa

\begin{figure}[tb]
\vskip 1 true cm
\includegraphics[width=17.0cm,keepaspectratio,angle=0,clip]{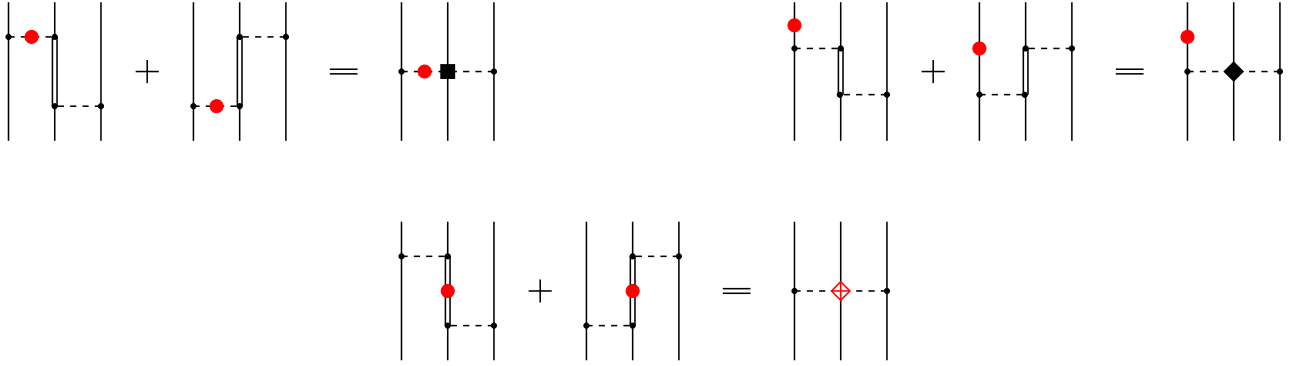}
    \caption{
         Diagrams in the $\Delta$-less EFT which reproduce the
         $\Delta$-contributions to the  leading
         isospin-breaking 3NF via resonance saturation of the LECs. Filled
         circles denote isospin-breaking pion, nucleon and delta mass
         shifts. Filled square and diamond refer to isospin-conserving
         vertices of dimension $\Delta_i = 1$ and $\Delta_i =2$. Crossed
         diamond denotes isospin-breaking strong and electromagnetic vertices
         of dimensions $\Delta_i = 4$ and $\Delta_i = 5$,
         respectively. For diagrams involving the insertion of the nucleon mass
         shift, only two representative graphs are shown. 
\label{fig2} 
 }
\end{figure}

It is instructive to understand how the obtained 3NF contributions are
reproduced in EFT without explicit deltas. The trivial $\Delta^{-1}$- and
$\Delta^{-2}$-dependence of the 3NF on the delta-nucleon mass splitting 
arising from the static propagator of the delta which enters Feynman diagrams (b-d) in
Fig.~\ref{fig1} ensures that the results are reproduced in the $\Delta$-less theory
by a \emph{finite} number of higher-order graphs (via resonance saturation of certain LECs).
This is depicted in Fig.~\ref{fig2}.
Here,  we have switched back to the original Lagrangian with the nucleon mass
shift which is more convenient to discuss resonance saturation. It should,
however, be understood that for Feynman diagrams involving the nucleon mass
shift, only irreducible contributions are taken into account.   
Consider first the CSC 3NF proportional to $\delta
M_\pi^2$ in Eq.~(\ref{res_mom}). As expected, this contribution is exactly reproduced  
in the $\Delta$-less theory by an appropriate shift in the LECs $c_3$
and $c_4$ \cite{Bernard:1996gq}, $c_3 = - 2 c_4 = -{4} h_A^2/(9 \Delta )$,
which in this case is the subleading\footnote{Numerically, however, it
 is expected to give the strongest isospin-breaking 3NF
 \cite{Epelbaum:2004xf}, see also the discussion in the next paragraph.} effect (i.e. order
$\nu = 5$). 
The remaining terms
in  Eq.~(\ref{res_mom}) are proportional to $\Delta^{-2}$ which means that
they arise in the $\Delta$-less
theory only at sub-subleading order $\nu = 6$. This is consistent with the absence of
CSB contributions proportional to $c_{3,4} \, \delta m_N$
at order $\nu =5$, see \cite{Epelbaum:2004xf,Epelbaum:2004qe}. It is easy to verify
that terms in Eq.~(\ref{res_mom}) proportional to $\delta m_N$
are indeed reproduced in the $\Delta$-less theory by $\Delta$-saturation of the
isospin-conserving sub-subleading pion-nucleon vertices (a complete list of
these terms in the Lagrangian can be found in Ref.~\cite{Fettes:1998ud}) while
the ones proportional to $\delta m_\Delta^1$
and  $\delta m_\Delta^2$ arise from resonance saturation of the
sub-subleading strong and electromagnetic isospin-breaking pion-nucleon
vertices, see Fig.\ref{fig2}.  

To get a rough idea about the size of the isospin breaking  
3NF contributions to e.g.~the 3N  binding energy,
one can look at the strength of the corresponding $r$-space potentials in 
Eq.~(\ref{r_space}). For the CSC terms $\propto \delta
M_\pi^2$ one gets $\delta M_\pi^2  g_A^2 h_A^2 M_\pi^4/(144 \pi^2 F_\pi^4
\Delta) \sim 50$ keV. Numerically, this is expected to be the biggest
isospin-breaking 3NF effect. Notice that in the theory without explicit deltas,
this contribution is shifted to the  subleading order $\nu = 5$
\cite{Epelbaum:2004xf}. The strength of
the remaining CSC 3NF contribution $\propto \delta m_\Delta^2$ is much
smaller, $| \delta m_\Delta^2 | \,   g_A^2 h_A^2 M_\pi^6/(384 \pi^2 F_\pi^4
\Delta^2) \sim 1.5$ keV (here we use the central value for 
$\delta m_\Delta^2 = -1.7$ MeV from Eq.~(\ref{fit1})). The estimated size of
the CSB 3NF in Eq.~(\ref{r_space}) using $\delta m_\Delta^1 = -5.3$ MeV is 
\beq
|\delta m_\Delta^1 - 3 \delta m_N | \,  \frac{g_A^2 h_A^2 M_\pi^6}{432 \pi^2 F_\pi^4
\Delta^2} \sim 3~{\rm keV}, 
\eeq
which is comparable to the typical size of the
leading CSB 3NF obtained in Ref.~\cite{Epelbaum:2004xf} and
based on EFT without explicit deltas, $\delta m_N
g_A^4 M_\pi^4/(256 \pi^2 F_\pi^4) \sim 7$ keV.  We further emphasize that the
CSB 3NF in Eq.~(\ref{res_mom}) vanishes exactly if one adopts the quark model
relation (\ref{furtherinput}).


\section*{Acknowledgments}

The work of E.E. and H.K. was supported in parts by funds provided from the 
Helmholtz Association to the young investigator group  
``Few-Nucleon Systems in Chiral Effective Field Theory'' (grant  VH-NG-222)
and through the virtual institute ``Spin and strong QCD'' (grant VH-VI-231). 
This work was further supported by the DFG (SFB/TR 16 ``Subnuclear Structure
of Matter'') and by the EU Integrated Infrastructure Initiative Hadron
Physics Project under contract number RII3-CT-2004-506078.


\begin{thebibliography}{10}
\bibitem{FuMi}
J.-I.~Fujita and H.~Miyazawa,  
Prog. Theo. Phys. {\bf 17}, 360 (1957).

\bibitem{TM1}
B.~H.~J.~McKellar and R.~Rajaraman,
Phys. Rev. Lett. {\bf 21}, 450 (1968).

\bibitem{TM2}
S.~A.~Coon, M.~D.~Scadron and B.~R.~Barrett,
Nucl. Phys. A {\bf 242}, 467 (1975).

\bibitem{Brazil}
H.~T.~Coelho {\em et al.}, Phys. Rev. C {\bf 28}, 1812 (1983).

\bibitem{Urbana1}
B.~S.~Pudliner {\em et al.}, Phys. Rev. C {\bf 56}, 1720 (1997).

\bibitem{Urbana2}
S.~C.~Pieper  {\em et al.}, Phys. Rev. C {\bf 64}, 014001 (2001).

\bibitem{GloeckleRept}
W.~Gl\"ockle {\em et al.}, Phys. Rept. {\bf 274}, 107 (1996).

\bibitem{KalantarNayestanaki:2007zi}
N.~Kalantar-Nayestanaki and E.~Epelbaum, Nucl. Phys. News {\bf 17}, 22 (2007),
[arXiv:nucl-th/0703089].



\bibitem{Weinberg:1991um}
S.~Weinberg,
\newblock Nucl. Phys. {\bf B363}, 3 (1991).

\bibitem{Hemmert:1997ye}
T.~R. Hemmert, B.~R. Holstein, and J.~Kambor,
\newblock J. Phys. {\bf G 24}, 1831 (1998), [arXiv:hep-ph/9712496].

\bibitem{Krebs:2007rh}
H.~Krebs, E.~Epelbaum, and U.-G. Mei{\ss}ner,
\newblock Eur. Phys. J. {\bf A32}, 127 (2007), [arXiv:nucl-th/0703087].

\bibitem{Tiburzi:2005na}
 B.~C.~Tiburzi and A.~Walker-Loud,
\newblock Nucl.\ Phys.\  A {\bf 764}, 274 (2006),
[arXiv:hep-lat/0501018].

\bibitem{Fettes:2000bb}
N.~Fettes and U.-G. Mei{\ss}ner,
\newblock Nucl. Phys. {\bf A679}, 629 (2001), [arXiv:hep-ph/0006299].

\bibitem{vanKolck:1994yi}
U.~van Kolck,
\newblock Phys. Rev. {\bf C49}, 2932 (1994).

\bibitem{Epelbaum:2005pn}
E.~Epelbaum,
\newblock Prog. Part. Nucl. Phys. {\bf 57}, 654 (2006), [arXiv:nucl-th/0509032].

\bibitem{Epelbaum:2002vt}
E.~Epelbaum {\em et~al.},
\newblock Phys. Rev. {\bf C66}, 064001 (2002), [arXiv:nucl-th/0208023].

\bibitem{Bernard:1996gq}
V.~Bernard, N.~Kaiser, and U.-G. Mei{\ss}ner,
\newblock Nucl. Phys. {\bf A615}, 483 (1997), [arXiv:hep-ph/9611253].

\bibitem{Friar:1998zt}
J.~L. Friar, D.~Huber, and U.~van Kolck,
\newblock Phys. Rev. {\bf C59}, 53 (1999), [arXiv:nucl-th/9809065].

\bibitem{Pandharipande:2005sx}
V.~R. Pandharipande, D.~R. Phillips, and U.~van Kolck,
\newblock Phys. Rev. {\bf C71}, 064002 (2005), [arXiv:nucl-th/0501061].

\bibitem{Epelbaum:2005fd}
E.~Epelbaum and U.-G. Mei{\ss}ner,
\newblock Phys. Rev. {\bf C72}, 044001 (2005), [arXiv:nucl-th/0502052].

\bibitem{Epelbaum:2004xf}
E.~Epelbaum, U.-G. Mei{\ss}ner, and J.~E. Palomar,
\newblock Phys. Rev. {\bf C71}, 024001 (2005), [arXiv:nucl-th/0407037].

\bibitem{Friar:1999zr}
J.~L. Friar and U.~van Kolck,
\newblock Phys. Rev. {\bf C60}, 034006 (1999), [arXiv:nucl-th/9906048].

\bibitem{Friar:2003yv}
J.~L. Friar, U.~van Kolck, G.~L. Payne, and S.~A. Coon,
\newblock Phys. Rev. {\bf C68}, 024003 (2003), [arXiv:nucl-th/0303058].

\bibitem{Friar:2004rg}
J.~L. Friar, G.~L. Payne, and U.~van Kolck,
\newblock Phys. Rev. {\bf C71}, 024003 (2005), [arXiv:nucl-th/0408033].

\bibitem{Friar:2004ca}
J.~L. Friar, U.~van Kolck, M.~C.~M. Rentmeester, and R.~G.~E. Timmermans,
\newblock Phys. Rev. {\bf C70}, 044001 (2004), [arXiv:nucl-th/0406026].

\bibitem{Urech:1994hd}
R.~Urech,
\newblock Nucl. Phys. {\bf B433}, 234 (1995), [arXiv:hep-ph/9405341].

\bibitem{Meissner:1997ii}
U.-G. Mei{\ss}ner and S.~Steininger,
\newblock Phys. Lett. {\bf B419}, 403 (1998), [arXiv:hep-ph/9709453].

\bibitem{Muller:1999ww}
G.~M{\"u}ller and U.-G. Mei{\ss}ner,
\newblock Nucl. Phys. {\bf B556}, 265 (1999), [arXiv:hep-ph/9903375].

\bibitem{Gasser:2002am}
J.~Gasser, M.~A. Ivanov, E.~Lipartia, M.~Moj\v{z}i\v{s}, and A.~Rusetsky,
\newblock Eur. Phys. J. {\bf C26}, 13 (2002), [arXiv:hep-ph/0206068].

\bibitem{Bernard:2005fy}
V.~Bernard, T.~R. Hemmert, and U.-G. Mei{\ss}ner,
\newblock Phys. Lett. {\bf B622}, 141 (2005), [arXiv:hep-lat/0503022].

\bibitem{Bernard:2003rp}
V.~Bernard, T.~R. Hemmert, and U.-G. Mei{\ss}ner,
\newblock Nucl. Phys. {\bf A732}, 149 (2004), [arXiv:hep-ph/0307115].

\bibitem{Rubinstein:1967aa}
H.~R. Rubinstein, F.~Scheck, and R.~H. Sokolov,
\newblock Phys. Rev. {\bf 154}, 1608 (1967).

\bibitem{PDG}
Particle Data Group, see the website http://pdg.lbl.gov/.

\bibitem{Gridnev:2004mk}
A.~B. Gridnev, I.~Horn, W.~J. Briscoe, and I.~I. Strakovsky,
\newblock Phys. Atom. Nucl. {\bf 69}, 1542 (2006), [arXiv:hep-ph/0408192].

\bibitem{Abaev:1995cx}
V.~V. Abaev and S.~P. Kruglov,
\newblock Z. Phys. {\bf A352}, 85 (1995).

\bibitem{Koch:1980ay}
R.~Koch and E.~Pietarinen,
\newblock Nucl. Phys. {\bf A336}, 331 (1980).

\bibitem{Pedroni:1978it}
E.~Pedroni {\em et~al.},
\newblock Nucl. Phys. {\bf A300}, 321 (1978).

\bibitem{Bernicha:1995gg}
A.~Bernicha, G.~Lopez~Castro, and J.~Pestieau,
\newblock Nucl. Phys. {\bf A597}, 623 (1996), [arXiv:hep-ph/9508388].

\bibitem{Arndt:2006bf}
R.~A. Arndt, W.~J. Briscoe, I.~I. Strakovsky, and R.~L. Workman,
\newblock Phys. Rev. {\bf C74}, 045205 (2006), [arXiv:nucl-th/0605082].

\bibitem{Gasser:1982ap}
J.~Gasser and H.~Leutwyler,
\newblock Phys. Rept. {\bf 87}, 77 (1982).

\bibitem{Beane:2006fk}
S.~R.~Beane, K.~Orginos, and M.~Savage, 
\newblock Nucl. Phys. {\bf B768}, 38 (2007), [arXiv:hep-lat/0605014].

\bibitem{Coleman:1969sm}
S.~R. Coleman, J.~Wess, and B.~Zumino,
\newblock Phys. Rev. {\bf 177}, 2239 (1969).

\bibitem{Haag:1958vt}
R.~Haag,
\newblock Phys. Rev. {\bf 112}, 669 (1958).

\bibitem{Epelbaum:2004qe}
E.~Epelbaum,
\newblock AIP Conf. Proc. {\bf 768}, 174 (2005), [arXiv:nucl-th/0412003].

\bibitem{Fettes:1998ud}
N.~Fettes, U.-G. Mei{\ss}ner, and S.~Steininger,
\newblock Nucl. Phys. {\bf A640}, 199 (1998), [arXiv:hep-ph/9803266].

\end{thebibliography}

\end{document}